\documentclass[mreferee,onecolumn]{gji}
\usepackage{times}
\usepackage{graphicx}
\title[Precision of noise correlation interferometry ] { On the precision of noise correlation interferometry  }
  
\author[R. L. Weaver et al.]
  {Richard L Weaver$^1$, C\'eline Hadziioannou$^2$, Eric Larose$^{2}\thanks{eric.larose@ujf-grenoble.fr}$, and Michel Campillo$^2$ \\
$^1$  Department of Physics,  University of Illinois at Urbana-Champaign, 1110 W Green, Urbana, Illinois 61801, USA.\\
  $^2$ Institut des Sciences de la Terre (ISTerre), Universit\'e J. Fourier \& CNRS, BP 53,\\ 38041 Grenoble cedex 9, FRANCE.
  }
    
\date{\today}

\begin{document}

\label{firstpage}

\maketitle

\begin{summary}
 Long duration noisy-looking waveforms such as those obtained in randomly multiply scattering and reverberant media are complex; they resist direct interpretation. Nevertheless, such waveforms are sensitive to small changes in the source of the waves or in the medium in which they propagate. Monitoring such waveforms, whether obtained directly or obtained indirectly by noise correlation, is emerging as a technique for detecting changes in media. Interpretation of changes is in principle problematic; it is not always clear whether a change is due to sources or to the medium. Of particular interest is the detection of small changes in propagation speeds. An expression is derived here for the apparent, but illusory, waveform dilation due to a change of source. The expression permits changes in waveforms due to changes in wavespeed to be distinguished with high precision from changes due to other reasons. The theory is successfully compared with analysis of a laboratory ultrasonic data set and a seismic data set from Parkfield California. 
\end{summary}


\section{Introduction}

The technique proposed in the 1980's \cite{poupinet1984} and later called "Coda wave interferometry" \cite{snieder2002} compares coda waveforms from multiply scattered waves obtained under different circumstances or on different dates and detects changes in a medium.  A multiply scattered wave can resist detailed interpretation, but for purposes of monitoring one may not need to interpret the waveform: it is sufficient to notice changes. Coda wave interferometry was first suggested for seismic waves but has also been applied in laboratory ultrasonics \cite{weaver2000,gorin2006,lobkis2008,derosny2001,lu2005}. In many such cases the change is due to a uniform change of temperature, and thus a uniform change in wave velocity.   To detect such changes, Weaver \& Lobkis \shortcite{weaver2000} constructed a dilation correlation coefficient between waveforms $\phi_1$ and $\phi_2$.

\begin{equation}
X(\epsilon) = \int \frac{ \phi_1(t) \phi_2 (t(1+\epsilon))dt}{\sqrt{\int \phi_1^2 (t) dt \int \phi_2^2 (t(1+\epsilon)) dt}}
\label{eq:corrcoef}
\end{equation}

$X$ takes on a value of unity at $\epsilon= 0$ if the two waveforms are identical. 
It will reach a value of unity at some characteristic value of $\epsilon$ if the two waveforms differ only by some temporal dilation. 
The estimated degree of dilation between two waveforms is taken to be the value of $\epsilon$ at which $X$ is maximum. 
$X$ reaches a maximum of less than unity if the waveforms differ by more than dilation alone. 
Therefore, the value of $X$ at its maximum, if it is less than unity, may be interpreted as a measure of the distortion between the waveforms.

An alternative formulation is Poupinet's doublet method \cite{poupinet1984}, which breaks $\phi_1$ and $\phi_2$ into a series of short time windows at several distinct times $t$, and determines the apparent shift $\delta t$ between them by examining conventional cross spectrum. $\delta t$ as a function of $t$, and in particular its slope $\delta t/t$ reveals a change in the medium.  Poupinet developed the doublet method in which seismic signals from repeated seismic events could be compared to infer changes in the earth \cite{poupinet1984}.   Song \& Richards (1996) and Zhang et al. (1985)  used this to show that certain earth crossing rays were shifted and distorted compared to versions some years earlier, indicating a relative rotation between the earth and its core.

The extensive literature in recent years on correlations of diffuse acoustic noise has reported theory and measurements in support of the notion that such correlations are essentially equal to the acoustic response that one would have at one receiver were there a source at the other \cite{lobkis2001,weaver2004,derode2003a,snieder2004,roux2005,gouedard2008a,tsai2009}.  More technically, what is recovered is the Green's function as filtered into the frequency band of the noise, whitened and symmetrized in time. Two different kinds of records can be correlated. Sometimes it is coda that is correlated \cite{campillo2003}. Coda waves consist in a long duration random looking signals that follows the main arrivals from a strong seismic source; coda waves are due to single and/or multiple scattering.   More commonly the diffuse noise is due to ambient seismic waves from continuously acting sources such as human activity or ocean storms.   Much recent literature reports constructions of the earth's seismic response between two seismograph stations, without the use of controlled sources, and without waiting for a seismic event.   Tomographic maps of seismic velocity with unprecedented resolution have been obtained  \cite{shapiro2005,sabra2005a}. The technique has even been applied on the moon \cite{larose2005}.  Very commonly, the noise which is correlated is incompletely equipartitioned, such that the resulting correlation waveforms do not precisely correspond to the Green's function.  A difference between noise correlation and Green's function can also be observed when one has not averaged enough raw data; the correlation may not have yet converged. Theoretical and applied work is ongoing in attempts to understand and correct for systematic errors due to these effects \cite{weaver2009,froment2010}. Nevertheless, Hadziioannou et al. (2009) demonstrated that it is not necessary to reconstruct the Green's function to use correlations for monitoring purposes.

These two approaches have been combined into what may be termed {\it noise-correlation interferometry} \cite{sabra2006,sens2006,brenguier2008a,brenguier2008b} in which correlations of seismic noise taken in different circumstances are compared. 
The correlations may have been obtained from different samples of ambient noise, perhaps on different dates, or from the codas of different events. The correlations are of course never identical; they are often very different. One reason for a difference is that the source of the noise may be different (yet if the correlation has converged to the local Green's function, a change of noise source ought to have little effect). Continuous seismic sources can move and strengthen or weaken as weather changes at sea.  It may also be that the correlation has not fully converged ({\it ie}, insufficient averaging has been done).  A third possibility is that the local mechanical or acoustic environment may have evolved, in particular, the local wave speed(s) may have changed. It is this possibility that is of particular interest, as changes in seismic velocities are associated with relaxations after major seismic events \cite{brenguier2008b}. In some cases changes in seismic velocity can be used to predict volcanic eruptions \cite{brenguier2008a}. Therefore, it is of great interest to be able to discern whether a change is due to a change in local environment or to a change in the character of the noise. The latter possibility is of some interest; the former is of great interest.

Our purpose here is to evaluate the precision with which wave speed changes can be evaluated. To do this we consider the case in which the two waveforms $\phi_1(t)$ and $\phi_2(t)$ differ only by noise so that the actual relative dilation without noise, is zero. We then ask for the apparent  (non-zero in general) value of {$ \epsilon$} at which the corresponding $X$ in equation (\ref{eq:corrcoef}) achieves its maximum.  The next section calculates the root mean square $\left< \varepsilon^2 \right>$ of this apparent, and erroneous, relative dilation.  The subsequent sections compare this prediction with experiment.

\section{Dilation Correlation Coefficient}

Here we examine the apparent waveform-dilation between two nominally identical signals. Theoretically, one ought to infer a relative dilation $\epsilon$ of zero, however, noise can corrupt the inference. 
Key to the following analysis is an understanding that the signals being discussed are like coda, in that they are statistically stationary with durations long compared to an inverse bandwidth.  We take the two waveforms to have an identical part $\psi(t)$, and to differ by noise $2 \mu \chi (t)$.   In the limit $\mu \rightarrow  0$, the waveforms become identical and have no relative dilation.  If $\mu \neq 0$, there will be an apparent, but actually meaningless, temporal dilation between them.  We wish to estimate this erroneous apparent relative dilation, and to identify any signatures that could be used to alert to the possibility of error.    Note that the common part $\psi$  of the signals need not be the local Green's functions.

We split the difference between these two waveforms $\phi_1$ and $\phi_2$, and define two signals $\psi$ and $\chi$; 

\begin{equation}
\phi_{1,2} = \psi (t) \pm \mu \chi (t)
\label{eq:signals}
\end{equation}

The waveform dilation-correlation coefficient  (\ref{eq:corrcoef}) between them is

\begin{eqnarray}
X(\epsilon,\mu) = \frac{\int \phi_1 (t(1+\epsilon /2 )) ~ \phi_2 (t(1-\epsilon /2)) ~ dt}{ \sqrt{ \int \phi_1^2 (t(1+\epsilon /2)) ~ dt ~ \int \phi_2^2 (t(1-\epsilon/2))~dt} } \nonumber\\
= \sqrt{1-\epsilon^2 /4} \frac{ \int \left[ \psi (t(1+\epsilon/2)) + \mu \chi (t(1+\epsilon /2)) \right] ~ \left[ \psi (t(1-\epsilon/2)) - \mu \chi (t(1-\epsilon/2)) \right] ~ dt }{ \sqrt{ \left[ \int (\psi^2 + \mu^2 \chi^2 ) ~dt \right]^2 - 4 \mu^2 \left[ \int \chi \psi ~dt \right]^2   }   } \\
=\sqrt{1-\epsilon^2 /4} ~ \frac{ N(\epsilon,\mu)}{D(\mu)} \nonumber
\label{eq:fraction}
\end{eqnarray}
with $N$ and $D$ defined  as, respectively, the numerator and the denominator of $X$.  The integrations are typically taken over a finite time-window with tapered edges. We make the approximation that the change of variable $t(1+\varepsilon/2)\to t$ and $t(1-\varepsilon/2)\to t$ in the denominator only leaves a prefactor $\sqrt{1-\varepsilon^2/4}$. 

The value of $\epsilon$ at which $X$ achieves its maximum is the practitioner's estimate of the dilation between the waveforms $\phi_1$ and $\phi_2$.  It occurs at $\epsilon$ such that $\partial X /\partial \epsilon = 0$, or,

\begin{equation}
0 = \sqrt{ 1- \epsilon^2/4} ~ D(\mu) \frac{\partial X(\epsilon,\mu) }{ \partial \epsilon } = -\frac{\epsilon ~ N(\epsilon,\mu)}{ 4}+(1-\epsilon ^2 /4)  \frac{\partial N(\epsilon,\mu)}{\partial \epsilon}.
\label{eq:isZero}
\end{equation}

If the phase shift due to time dilation is much less than one oscillation, which implies $t \omega \epsilon \ll 1$ for all times $t$ and frequencies $\omega$ of interest, it suffices to expand $N(\epsilon,\mu)$ through only the second power of $\epsilon$:

\begin{eqnarray}
N(\epsilon,\mu) = \int \left[ \psi (t) + \frac{t\epsilon}{2} \dot{\psi}(t) + \frac{t^2\epsilon^2}{8} \ddot{\psi}(t) + \mu \chi (t) + \frac{\mu \epsilon t}{2} \dot{\chi}(t) + \frac{\mu t^2 \epsilon^2}{8} \ddot{\chi}(t) \right] \nonumber \\
 \times ~ \left[ \psi (t) - \frac{t \epsilon}{2} \dot{\psi}(t) + \frac{t^2 \epsilon^2}{8} \ddot{\psi}(t) - \mu \chi (t) + \frac{\mu \epsilon t}{2} \dot{\chi}(t) - \frac{\mu t^2 \epsilon^2}{8} \ddot{\chi}(t) \right] ~ dt.
\end{eqnarray}

On collecting terms in $N(\epsilon,\mu)$ that are linear and quadratic in $\epsilon$ obtains:

\begin{eqnarray}
N(\epsilon,\mu) & \sim & \int \left[ \psi(t)^2 - \mu ^2 \chi (t)^2 \right]  dt + \int \left[ \frac{t\epsilon}{2}\dot{\psi}(t) +  \frac{\mu \epsilon t}{2} \dot{\chi}(t) \right] ~ \left[ \psi(t) - \mu \chi(t) \right] dt \nonumber \\
&+& \int \left[ - \frac{t\epsilon}{2}  \dot{\psi}(t) + \frac{\mu \epsilon t}{2} \dot{\chi}(t) \right]~ \left[  \psi(t) + \mu \chi(t)  \right] dt \nonumber \\
&+& \int \left[ \frac{t^2 \epsilon^2}{8}  \ddot{\psi}(t) + \frac{t^2 \mu \epsilon^2}{8} \ddot{\chi}(t) \right] ~ \left[ \psi(t) - \mu \chi(t) \right] dt \nonumber \\
&+& \int \left[ \frac{t^2 \epsilon^2}{8}  \ddot{\psi}(t) - \frac{t^2 \mu \epsilon^2}{8} \ddot{\chi}(t) \right] ~ \left[ \psi(t) + \mu \chi(t) \right] dt \nonumber \\
&+& \int \left[  \frac{t\epsilon}{2}  \dot{\psi}(t) + \frac{\mu \epsilon t}{2} \dot{\chi}(t) \right] \left[ - \frac{t\epsilon}{2}  \dot{\psi}(t) + \frac{\mu \epsilon t}{2} \dot{\chi}(t) \right] dt
\end{eqnarray}

\begin{eqnarray}
&=& \int \left[ \psi (t)^2 -  \mu ^2 \chi(t)^2 \right] dt + \epsilon \int t \mu \left[ (\dot{\chi}(t)\psi(t) - \chi(t) \dot{\psi}(t) \right] dt \nonumber \\
&& + \frac{\epsilon^2}{4} \int t^2 \left[ \ddot{\psi}(t)\psi(t) - \mu^2 \ddot{\chi}(t)\chi(t) \right] dt \\
&& - \frac{\epsilon^2}{4} \int t^2 \left[ \dot{\psi}(t)^2 - \mu^2 \dot{\chi}(t)^2 \right] dt.	\nonumber
\end{eqnarray}

The first term in $\epsilon^2$ may be integrated by parts.

\begin{eqnarray}
N(\epsilon,\mu)& \sim & \int \left[ \psi(t)^2 - \mu^2 \chi(t)^2 \right] dt + \epsilon \int t \mu \left[ \dot{\chi}(t) \psi (t) - \chi(t)\dot{\psi}(t) \right] dt	\nonumber \\
&&-\frac{1}{2} \epsilon^2 \int t^2 \left[ \dot{\psi}(t)^2 - \mu^2 \dot{\chi}(t)^2 \right] dt + \frac{1}{4} \epsilon ^2 r
\label{eq:intParts}
\end{eqnarray}
where the quantity $r$ is:

\begin{equation}
r=\int \left[ \psi (t)^2 - \mu^2 \chi(t)^2 \right] dt - t \left[ \psi(t)^2 - \mu^2 \chi(t)^2 \right]  -t^2 \left[ \psi(t) \dot{\psi}(t) - \mu^2 \chi(t) \dot{\chi}(t) \right], 
\end{equation}
whose expectation is zero and whose typical value is much less (by a factor of $t^2 \omega^2$) than the other coefficient of $\epsilon^2$ in Eq.~(\ref{eq:intParts}). For this reason we henceforth neglect it. 

So finally, neglecting high order of $\varepsilon$,

\begin{equation}
\partial N (\epsilon,\mu)/\partial \epsilon \sim \int \mu t \left[ \dot{\chi}(t) \psi(t) - \chi(t) \dot{\psi}(t) \right] dt - \epsilon \int t^2 \left[ \dot{\psi}(t)^2 - \mu^2 \dot{\chi}(t)^2 \right] dt.
\end{equation}

Equation~(\ref{eq:isZero}) is satisfied for:
\begin{equation}
\epsilon = n/d,
\label{eq:ND}
\end{equation}
with

\begin{eqnarray}
n & = & \mu \int t \left[ \dot{\chi}(t) \psi(t) - \chi (t)\dot{\psi}(t) \right] dt \nonumber \nonumber, \\
d & = & \int \left[ t^2 ( \dot{\psi}(t)^2 - \mu^2 \dot{\chi}(t)^2 ) \right] dt + \frac{1}{4} \int \left[ \psi (t)^2 - \mu^2 \chi(t)^2 \right] dt.  \nonumber
\end{eqnarray}

Equation~(\ref{eq:ND}) is an expression for the apparent dilation induced by the difference $2\mu \chi$ between the original waveforms $\phi_1$ and $\phi_2$. Given specific $\psi$ and $\chi$, one could evaluate it. It will be more useful, however, to obtain statistical estimates for the apparent dilation given assumptions about the envelopes and spectra of $\psi$ and $\chi$.   The numerator $n$ has expectation zero, as $\chi$ and $\psi$  are statistically unrelated.   Thus within the stated limit $\omega t \epsilon << 1$, differences $\phi_2 - \phi_1$ do not manifest as an apparent dilation and the expected dilation $\epsilon$ is zero. 

\section{Variance estimation and statistical error}
Given $\langle n \rangle =0$, one then seeks estimates for the root-mean-square of equation~(\ref{eq:ND}) in order to judge typical fluctuations around the expected zero.  These will be made based on assumptions that $\psi$ and $\chi$ are stationary, noise-like and Gaussian, with similar spectra, having central frequency $\omega_c$. $\psi$ and $\chi$ have the same duration, long compared to the inverse of $\omega_c$.  Without loss of generality it is also assumed that they have the same amplitudes $\langle \psi^2 \rangle = \langle \chi^2 \rangle = 1$. They are taken to extend from a start time $t_1$ to an end time $t_2$. Under these assumptions the denominator of (\ref{eq:ND}) is estimated as  :

\begin{equation}
d \approx (1 - \mu^2) \left[ \frac{1}{3} \omega_c^2 (t^3_2 - t^3_1) + \frac{1}{4} (t_2 - t_1) \right] \approx (1-\mu^2) \frac{1}{3} \omega_c^2 (t^3_2 - t^3_1).
\label{eq:denom}
\end{equation}
The square of the numerator of (\ref{eq:ND}) is

\begin{equation}
n^2 \approx \mu^2 \left[ \int\int t t' \left\{ \psi(t) \dot{\chi}(t) - \dot{\psi}(t) \chi(t) \right\} \left\{ \psi (t') \dot{\chi}(t') - \dot{\psi}(t') \chi(t') \right\} dt ~ dt' \right].
\label{eq:num1}
\end{equation}

On changing variables: $t+t' = 2\tau $, $t-t' = \xi$  and dropping the cross terms as having expectation zero, (\ref{eq:num1}) becomes:

\begin{eqnarray}
\langle n^2 \rangle  \approx   \mu^2 \int \left(\tau^2 - \frac{\xi^2}{4} \right) \left\{ \psi \left( \tau + \frac{\xi}{2} \right) \dot{\chi} \left( \tau + \frac{\xi}{2} \right) \psi \left( \tau - \frac{\xi}{2} \right) \dot{\chi} \left( \tau - \frac{\xi}{2} \right)  \right. \nonumber\\
 \left. +    \dot{\psi} \left( \tau + \frac{\xi}{2} \right) \chi \left( \tau + \frac{\xi}{2} \right)  \dot{\psi} \left( \tau - \frac{\xi}{2} \right) \chi \left( \tau - \frac{\xi}{2} \right)  \right\} d\tau ~ d\xi .
\end{eqnarray}

Auto-correlation functions may be defined as

\begin{equation}
\langle {\psi \left( \tau + \frac{\xi}{2} \right) \psi \left( \tau - \frac{\xi}{2} \right) \rangle = \langle \psi^2(\tau) }\rangle  R_{\psi}(\xi) = R(\xi),
\end{equation}
such that

\begin{equation}
\langle \dot{\psi} \left( \tau + \frac{\xi}{2} \right)  \dot{\psi} \left( \tau - \frac{\xi}{2} \right)  \rangle \approx \omega_c^2 \langle \psi^2(\tau) \rangle R_{\psi}(\xi) = \omega_c^2 R (\xi),
\end{equation}
with similar expressions for $\chi$. Then the expectation of the square of the numerator of (\ref{eq:ND}) is:

\begin{equation}
\langle n^2 \rangle \approx 2 \mu^2 \left[ \int \left(\tau^2 - \frac{\xi^2}{4} \right) \omega_c^2 \; R^2(\xi) \; d\tau ~ d\xi \right] \approx 2 \; \mu^2 \; \omega_c^2  \left[ \int \tau^2 d\tau \right] \left[ \int R^2 (\xi) d\xi \right].
\label{eq:num2}
\end{equation}

The first integral is merely $(t_2^3 - t_1^3)/3$. The second requires knowing something of the spectra of $\psi$ and $\chi$, so we take these to be Gaussian and identical: $\sim \exp ( - (\omega - \omega_c)^2 T^2) + \exp ( - (\omega + \omega_c)^2 T^2)$. $T$ may be identified by noting that the -10 dB points are at $\omega_c$ plus or minus  $\ln10/T$. In this case $R$ is related to the inverse Fourier transform of the power spectrum: $R(\xi) = \cos (\omega_c \; \xi) \exp(\xi^2/4T^2)$.  Then the second integral in (\ref{eq:num2}) is identified as $T \sqrt{\pi/2}$.

Application of equations~(\ref{eq:ND}), (\ref{eq:denom}), (\ref{eq:num2}) requires that we also estimate the $\mu$. The quantity $\mu$ is related to the maximum of the waveform dilation-correlation coefficient

\begin{equation}
X(0,\mu) = \frac{N(0,\mu)}{D(\mu)} = \frac{  \int  \psi(t)^2 - \mu^2 \chi(t)^2  dt  }{ \sqrt{ \left[ \int  \psi^2 + \mu^2 \chi^2  dt \right] ^2 - 4\mu^2 \left[ \int \chi \; \psi ~ dt \right] ^2 }    }
\end{equation}
As $\chi$ and $\psi$ are statistically independent, one estimates the following relation between the maximum of the dilation correlation coefficient and the parameter $\mu$: 

\begin{equation}
X = \frac{ 1-\mu^2}{ 1 + \mu^2}
\end{equation}
Finally, the root mean square of the practitioner's (erroneous) estimate for the relative dilation between $\phi_1$ and $\phi_2$ is 

\begin{equation}
\rm{rms}  ~  \epsilon = \frac{\langle n^2 \rangle ^{1/2}}{d} = \frac{\sqrt{1-X^2}}{2X} \sqrt{ \frac{ 6 \sqrt{\frac{\pi}{2}}T } { \omega_c^2 \; (t_2^3 - t_1^3)} }
\label{eq:rms}
\end{equation}
We recall that $T$ is the inverse of the frequency bandwidth, $t_1$ and $t_2$ are begin and end time of the processed time-window in the coda, respectively, and $\omega_c$ is the central pulsation. This expression scales inversely with the duration of the correlation waveform in units of the period, and inversely with the square root of the duration in units of the inverse bandwidth.   In practice Eq.~(\ref{eq:rms}) can be very small.  The quantity $ \omega(t_2-t_1)$ represents the available time where coda waves are significantly larger than the noise; the duration of the waveform is in units of the period. The quantity $T$ is the amount of time for one bit of information to be delivered, and corresponds roughly to the time of the initial source \cite{derode1999}. Thus Eq.~(\ref{eq:rms}) can be recognized as scaling inversely the available time in the coda (this time is related to coda-Q), and inversely with the square root of the amount of information. It also may be recognized that small $X$ corresponding to waveforms $\phi_1$ and $\phi_2$ that are very different, permits the practitioners erroneous estimate of dilation to be large.  It may be that lengthening the considered time interval $t_2-t_1$ would increase the precision, however it could also diminish $X$: in principle there are trade-offs.

Application of Eq.~(\ref{eq:rms}) is straightforward. A practitioner's estimate of the relative dilation $\epsilon$ between two waveforms  $\phi_1$ and $\phi_2$ may be compared to Eq.~(\ref{eq:rms}).  Values in excess of Eq.~(\ref{eq:rms}) are consistent with the inference that the observed dilation is real. Changes in waveform source or other character should not generate apparent dilations in excess of Eq.~(\ref{eq:rms}). Furthermore, in absence of any actual dilation, estimates of $\epsilon$ of the order Eq.~(\ref{eq:rms}) will nevertheless be generated in practice. Such should be regarded as un-meaningful.

\section{Comparison with Experiment}

\begin{figure}
\vspace{1cm}
\includegraphics{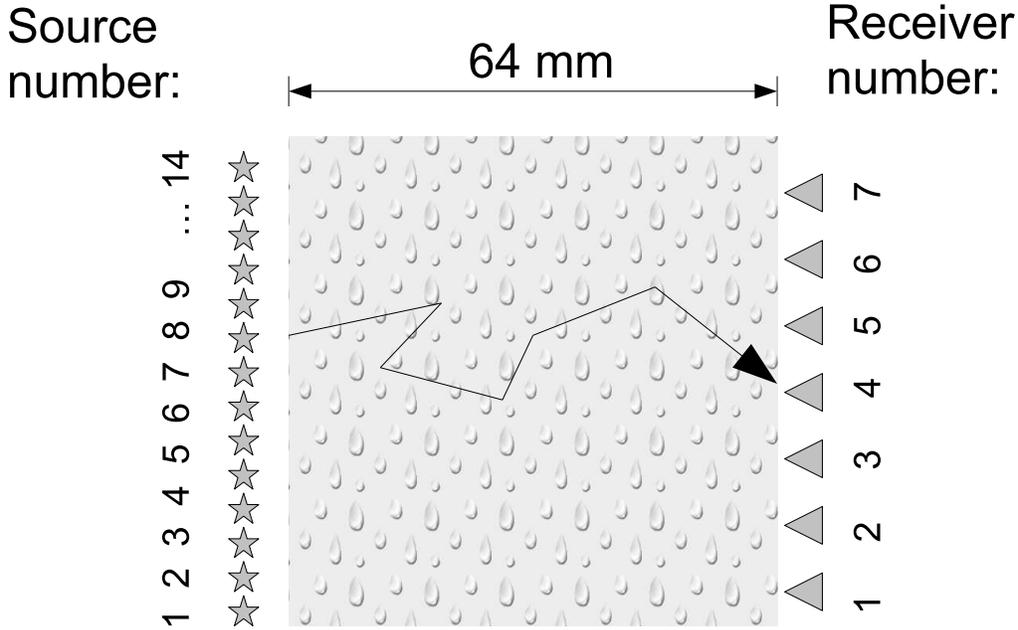}
\caption{ Experimental setup with ultrasound. An air-gel mixture mimics a multiple scattering medium. Coda waves sensed by the receivers (see Fig.~\ref{fig:diffSig}) are processed like ambient seismic noise : they are autocorrelated and compared from one date to another.}
\label{fig:exp}
\end{figure}

\begin{figure}
\vspace{1cm}
\includegraphics[width=10cm]{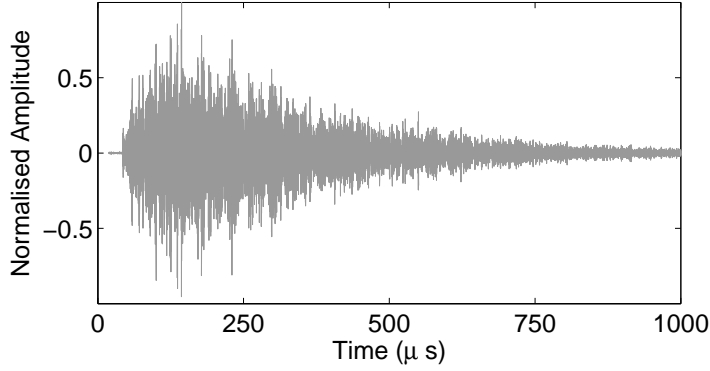}
\caption{ A typical signal $f_{sr}(t)$ in the gel resembles noise, under an envelope which is a solution of a diffusion equation.}
\label{fig:diffSig}
\end{figure}

\begin{figure}
\vspace{1cm}
\includegraphics[width=10cm]{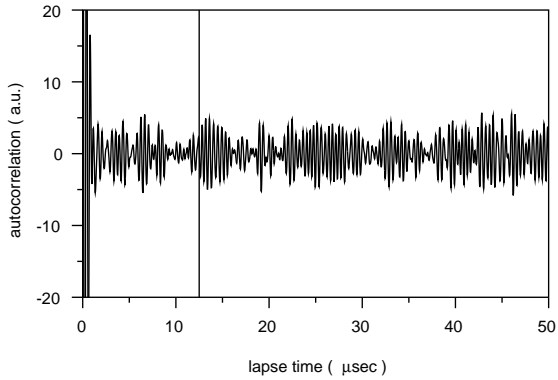}
\caption{A typical autocorrelation $g_{sr}$, of the signal from one of the sources to one of the
receivers. The interval from 12.5 to 50 $\mu$sec was selected for dilation coefficient evaluation. }
\label{fig:autocorr}
\end{figure}

The prediction Eq.~(\ref{eq:rms}) has been compared to waveform dilation measurements in a laboratory ultrasonic experiment \cite{hadziioannou2009}. For practical reasons, we mimicked here ambient noise correlation with diffuse waves correlation. Several piezoelectric sensors and sources were applied to a multiply scattering air-bubble filled gel (see Fig.~\ref{fig:exp}). Sources and receivers were placed on opposite sides, 64 mm apart. Multiple scattering was strong: received waveforms $f_{sr}(t)$ from sources $s$ to receivers $r$ were coda-like, with envelopes that resembled the solution of a diffusion equation (fig.~\ref{fig:diffSig}). The auto-correlation of each $f_{sr}(t)$ was windowed between lapse times of 12.5 to 50 $\mu$sec, to yield the waveforms which we call $g_{sr} (\tau)$ (see fig.~\ref{fig:autocorr}). Details of the experimental set-up are described in Hadziioannou et al (2009). The details are, however, unimportant here, as the present theory applies to any pair of coda-like waveforms $\phi_1$ and $\phi_2$. The typical $g_{sr}$ is stationary over this interval and has a power spectrum centered on 2.35  MHz with -10 dB points at 1.7 and  3.0 MHz.

The tables below are formed by maximizing the dilation correlation coefficient $X$ between sums $\phi = \sum_s g_{sr}$ over different sets of sources $\{ s \}$.  Note that the $\phi$ are not Green's functions $G_{rr}$, as the fields $f_{sr}(t)$ used to compose them were not fully equipartitioned. The excellent impedance match between the gel and the receivers prevented the field to be reflected back to the medium. The noise field thus lacked any components traveling from receiver to receiver. All tests were conducted at fixed temperature, the actual relative temporal dilation is therefore zero. Also, to mimic signals acquired at different date, we averaged correlations over different set of sources to eventually compare them. The addition of different sources results in an additional noise $\chi(t)$ in the correlations.
The goal of the test is to measure the dilation induced by the difference in waveform due to a different source distribution.

Autocorrelation waveforms, like that illustrated in figure~\ref{fig:autocorr}, in the interval from 12.5 to 50 $\mu$sec appear stationary. Thus we take $t_1$=12.5 $\mu$sec, $t_2$ = 50 $\mu$sec,  $\omega_c$ = 15 rad/$\mu$sec;  and $T$ = 0.56 $\mu$sec and conclude from (\ref{eq:rms}),

\begin{equation}
\rm{rms} ~ \epsilon = 4 \times 10 ^{-4} ~  \frac{\sqrt{1-X^2}}{2X}
\label{eq:errLab}
\end{equation}

Tables \ref{tb:11src} and \ref{tb:6src} show two case studies. In the first case, autocorrelations calculated from the signals at a receiver $r$, as produced by eleven distinct sources $s$, were summed over to generate the reference waveform $\phi_1 = \sum_s g_{sr}$.  
For each of three comparison waveforms $\phi_2$, the same sum was done, keeping the first ten sources unchanged. In order to deliberately change the waveform without dilation, the eleventh source is replaced  with sources number twelve, thirteen and fourteen respectively. 
This was repeated for each of seven receivers. 
In each case we compare three waveforms $\phi_2$ with the reference $\phi_1$ and evaluate $X(\epsilon)$. 
The table shows the maximum value of $X(\epsilon)$, and the value of $\epsilon$ that did this, for each of the 21 cases. 
For each  of the seven receivers we calculate the {\it rms} of these three $\epsilon$. 
If the only changes were to the source of the noise field, and not the medium, one would expect no dilation, or $\epsilon = 0$. 
Nevertheless, the differences in sources do generate apparent (feeble but noticeable) dilations $\sim\epsilon$. Theoretical and experimental $rms(\varepsilon)$ are of the same order of magnitude. Theory, especially in light of the approximate modeling of the spectrum, may be said to have done a good job predicting the fluctuations.

\begin{table}
\begin{tabular}{|c|c|c|c|c|c|c|c|}
\multicolumn{8}{c}{$X$ for  seven receivers and three different choices for the set of sources} \\
\hline
Sources 1 to 11 \& 12 & 0.9312 & 0.9169 & 0.8893 & 0.8458 & 0.8226 & 0.8852 & 0.8683 \\ 
Sources 1 to 11 \& 13 & 0.9394 & 0.8833 & 0.9083 & 0.8464 & 0.8872 & 0.8631 & 0.8928 \\
Sources 1 to 11 \& 14& 0.9458 & 0.9009 & 0.8730 & 0.7942 & 0.8322 & 0.8396 & 0.7979 \\
\hline
\multicolumn{8}{c}{}\\
\multicolumn{8}{c}{The dilation $\epsilon$ ($\times 10^{-3}$) as obtained by maximizing $X$ for each of these cases}\\
\hline
Sources 1 to 11 \& 12 &0.06 & 0.04 & -0.16 & 0.08 & -0.04 & -0.10 & 0.18 \\
Sources 1 to 11 \& 13 &-0.04 & -0.04 & -0.08 & 0.06 & -0.14 & -0.08 & 0.10 \\ 
Sources 1 to 11 \& 14 &-0.16 & 0.08 & -0.12 & 0.00 & -0.14 & -0.24 & -0.12 \\ 
\hline
\multicolumn{8}{c}{}\\
\multicolumn{8}{c}{Experimental root mean square dilation $\epsilon$ ($\times 10^{-3}$)}\\
\hline
all sets & 0.1013 & 0.0566 & 0.1244 & 0.0577 & 0.1166 & 0.1571 & 0.1376 \\
\hline
\multicolumn{8}{c}{}\\
\multicolumn{8}{c}{Theoretical root mean square ($\times 10^{-3}$) from Eq.~(\ref{eq:errLab})}\\
\hline
all sets & 0.07 & 0.09 & 0.09 & 0.12 & 0.11 & 0.10 & 0.11 \\ 
\hline
\end{tabular}
\caption{Comparison of best-fit waveform dilations $\epsilon$ with the predictions of equation (\ref{eq:errLab}).   A maximum value of $X$ and the $\epsilon$ at which that $X$ is maximum, are constructed for each of seven receivers (the seven columns) and the three choices for the set of sources described in the text (the three rows). The root mean square of those $\epsilon$ is compared with the predictions of theory. That $X$ is of order 90\% is consistent with one source in ten having changed.}
\label{tb:11src}
\end{table}

In the second study (Tab.~\ref{tb:6src}), four sources were held constant, and two were varied. Here the reference waveform was constructed from a sum over six sources $\sum_s g_{sr}$; each of the other three waveforms was constructed by replacing sources number five and six in that sum with two others. Again theory may be said to have done a good job: the $rms$  theoretical predictions accurately fit the actual experimental errors within 40\%. This means that Eq.~\ref{eq:rms} properly predicts the order of magnitude of the error.

\begin{table}
\begin{tabular}{|c|c|c|c|c|c|c|c|}
\multicolumn{8}{c}{$X$ for  seven receivers and three different choices for the set of sources} \\
\hline
Sources 1 to 4 \& 5-6& 0.6181 & 0.7864 & 0.7143 & 0.8400 & 0.7149 & 0.8458 & 0.7863 \\
Sources 1 to 4 \& 7-8 & 0.6359 & 0.7340 & 0.7011 & 0.8451 & 0.8020 & 0.8285 & 0.8194 \\
Sources 1 to 4 \& 9-10& 0.5948 & 0.7397 & 0.5837 & 0.8165 & 0.8294 & 0.8451 & 0.8745 \\
\hline
\multicolumn{8}{c}{}\\
\multicolumn{8}{c}{The dilation $\epsilon$ ($\times 10^{-3}$) as obtained by maximizing $X$ for each of these cases}\\
\hline
Sources 1 to 4 \& 5-6 & -0.0800 & 0.0400 & 0.4600 & -0.0200 & 0.1400 & 0.0400 & 0.1600 \\ 
Sources 1 to 4 \& 7-8 & -0.0200  & 0.0400 & 0.0400 & -0.3400 & -0.0800 & -0.1400 & 0.0800 \\
Sources 1 to 4 \& 9-10 & -0.4600 & 0.1600 & 0.5600 & -0.0400 & 0.0800 & -0.0400 & 0.0400 \\
\hline
\multicolumn{8}{c}{}\\
\multicolumn{8}{c}{Experimental root mean square dilation $\epsilon$ ($\times 10^{-3}$)}\\
\hline
all sets & 0.27 & 0.098 & 0.419 & 0.198 & 0.104 & 0.087 & 0.106 \\
\hline 
\multicolumn{8}{c}{}\\
\multicolumn{8}{c}{Theoretical root mean square ($\times 10^{-3}$) from Eq.~(\ref{eq:errLab})}\\
\hline
all sets & 0.19 & 0.14 & 0.17 & 0.12 & 0.14 & 0.12 & 0.11 \\
\hline
\end{tabular}
\caption{As in table~\ref{tb:11src}, but for sources that differ by more: fewer sources are kept fixed (four) and more sources are changing (two). This results in as maller values of $X$ and a larger value of error (uncertainty).}
\label{tb:6src}
\end{table}

\section{Comparison with Seismic Data from Parkfield}

\begin{figure}
\vspace{1cm}
\includegraphics[width=10cm]{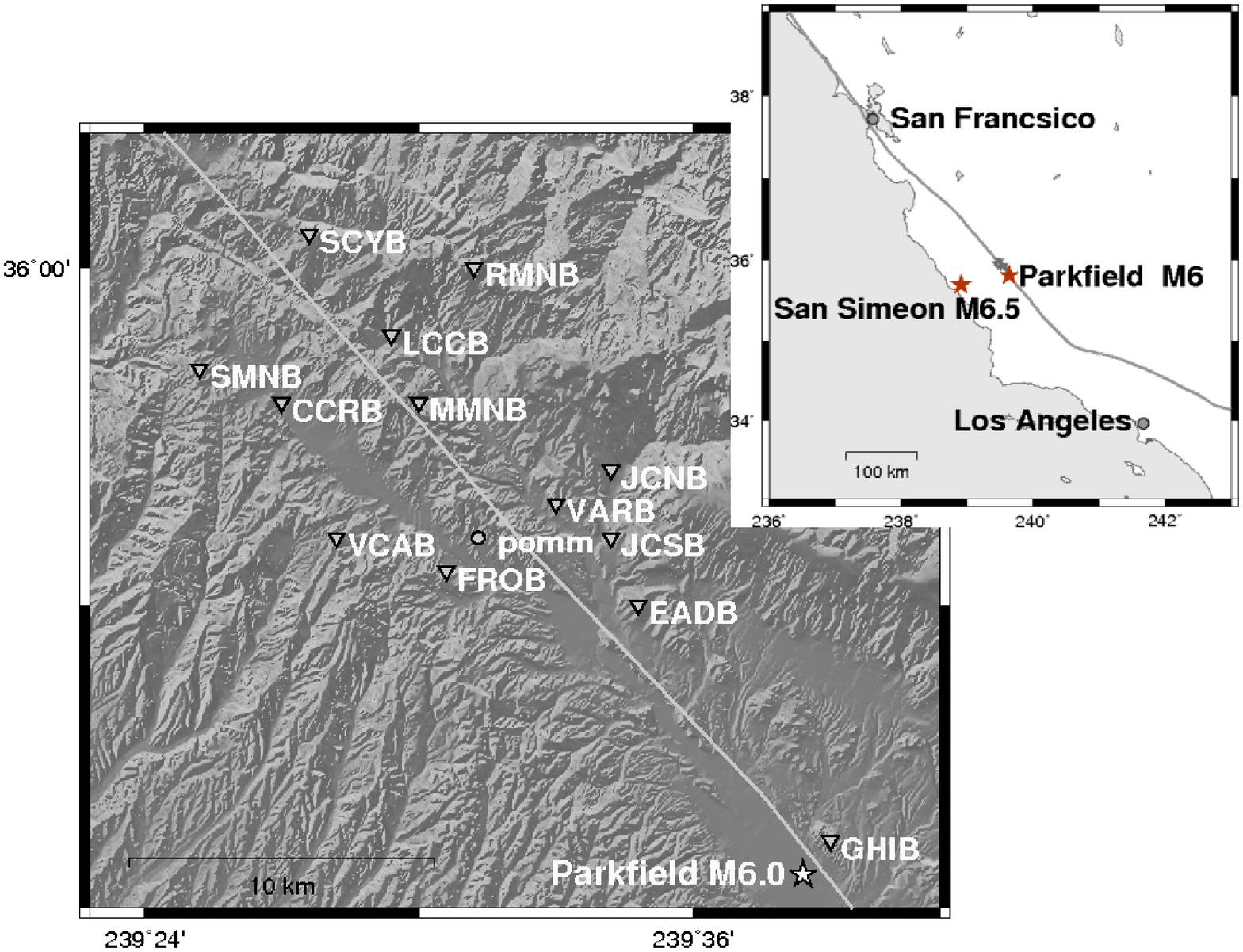}
\caption{Map of the seismic stations used in our study. They are part of the High Resolution Seismic Network operated by the Berkeley Seismological Laboratory.}
\label{fig:map}
\end{figure}

\begin{figure}
\vspace{1cm}
\includegraphics[width=10cm]{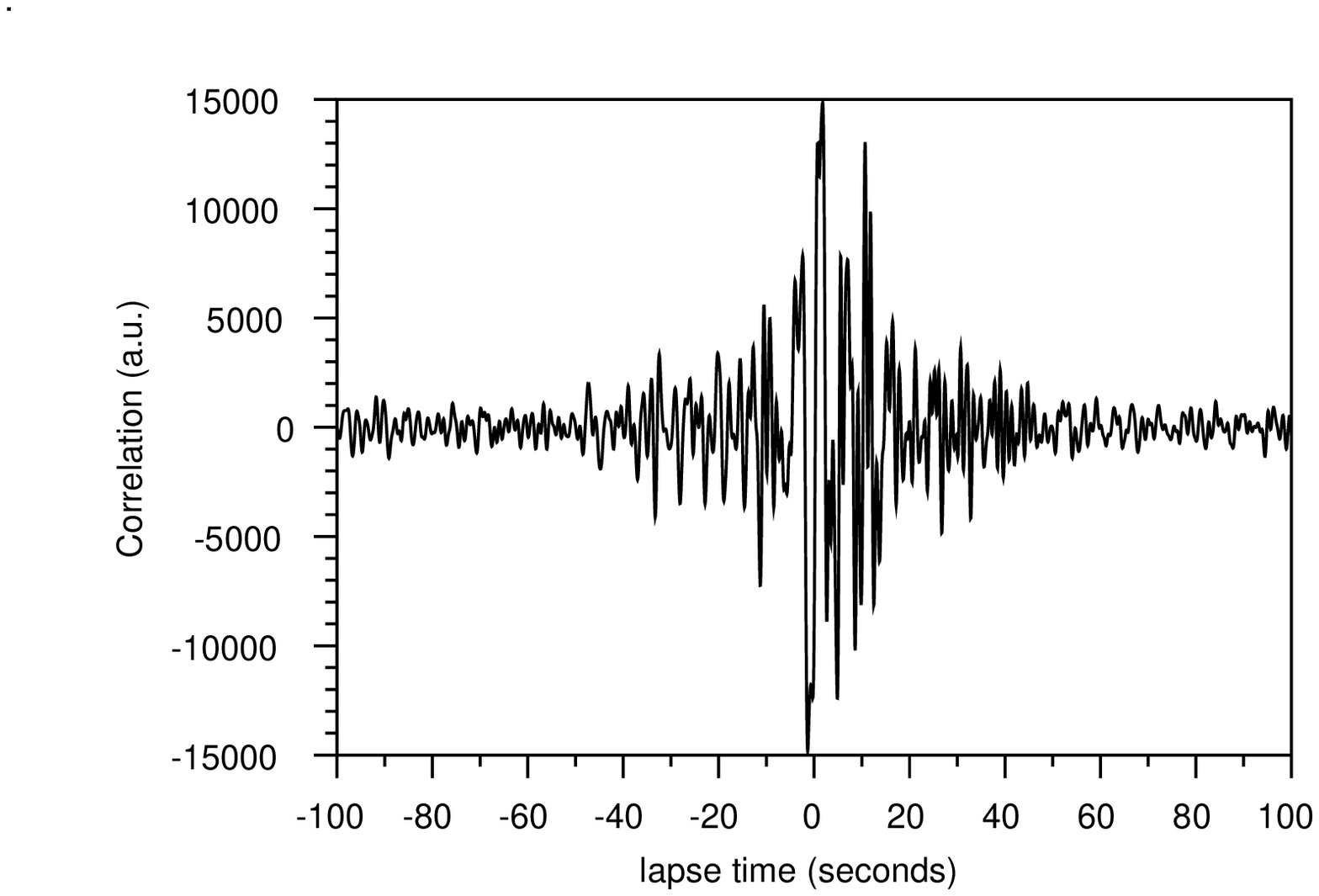}
\caption{A typical daily correlation waveform from the Parkfield data set. Dilations were
constructed by comparing waveforms like this as windowed from Ð50 to +50 seconds, with the
direct signal from Ð20 to +20 omitted.}
\label{fig:dailycorr}
\end{figure}

We also analyze data from seismic measurements near Parkfield, California. Brenguier et al. (2008) showed that correlation waveforms obtained from ambient seismic noise over a period of five years from 2002 to 2007 changed in a manner consistent with a decrease of the seismic velocity after the earthquake of 2004 (Fig.~\ref{fig:map}). This decreased velocity then relaxed like $\log(t)$ after the earthquake. While they used the doublet technique, we have re-analyzed their data using the dilation coefficient (see Eq.~\ref{eq:corrcoef}). 
For each of 78 receiver pairs, we compared the 1550-day average correlation waveform with the correlation waveforms constructed from each of 1546 overlapping 5-day segments. The whitening operation before correlation ensures that the spectrum of the correlations is constant. Note that direct arrivals are never processed. A representative correlation waveform is shown in figure~\ref{fig:dailycorr}. Each such waveform was windowed between -50 and -20 seconds, and again from 20 to 50 seconds (thus excluding direct Rayleigh arrivals and emphasizing the multiply scattered diffuse part of the signal for which the theory was developed). As in the previous section, the details of the measurements are available elsewhere \cite{brenguier2008b} but are unimportant for the present purposes.   An $X$ and an $\epsilon$ were deduced for each day. Power spectra were centered on 0.5 Hz, with -10dB shoulders at 0.1 and 0.9 Hz. These numbers permit the evaluation of (\ref{eq:rms}):

\begin{equation}
\rm{rms} ~ \epsilon = 2,4 \times 10 ^ {-3} ~ \frac{\sqrt{1-X^2}}{2X}
\label{eq:PFerr}
\end{equation}

\begin{figure}
\vspace{1cm}
\includegraphics[width=10cm]{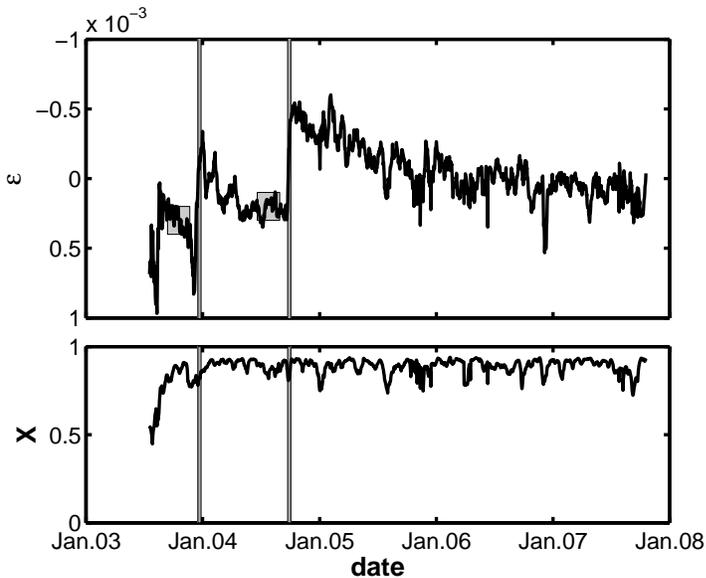}
\caption{Top: Dilation $\epsilon$ averaged over 78 receiver pairs, using a 5-day sliding window. The grey squares indicate the 70-day reference windows. The best fit $\epsilon$ varied weakly and stochastically over this period, with two notable jumps, after December 22, 2003 and after the Parkfield earthquake on September 28, 2004. The latter jump was followed by a slow recovery. Fluctuations have an rms strength of about $10^{-4}$. Bottom: Dilation coefficient $X$. the maximum value of the dilation coefficient $X$, averaged over 78 receiver pairs. }
\label{fig:PFdvv}
\end{figure}

Figures~\ref{fig:PFdvv} (top and bottom) show the mean (over the 78 receiver pairs) values of $X$ and $\epsilon$ between each of the 1546 overlapping 5-day correlation waveforms, $\phi_1$, and the correlation waveform $\phi_2$ as obtained by averaging over the entire 5 year period.   Except for the two events on days 152 and 437, and the slow relaxation after the latter, the dilation appears constant, with daily random fluctuations of order $10^{-4}$. A correlation coefficient $X$ of 0.8 predicts a $rms$ fluctuation of $10^{-3}$ (Eq.~(\ref{eq:PFerr})). On averaging over 78 pairs, this prediction is reduced by a factor $\sqrt{78}$, to $1.1 \times 10^{-4}$, consistent with the observed fluctuations in $\epsilon$. In light of the approximations, in particular that of modeling the spectrum as Gaussian and the waveform as stationary, we count this as excellent agreement.

The discontinuities in $\epsilon$ at December 22, 2003 and September 28, 2004 are of particular interest. The latter is coincident with the Parkfield earthquake. 
Jumps in dilation on those dates by $ \sim 0.8  \times 10^{-3}$ were interpreted \cite{brenguier2008b} as decrease of local seismic wavespeed. But one might wish to entertain the hypothesis that these jumps are due to a change in the source of the noise. 
To examine the question, we evaluated $X$ and $\epsilon$ using correlation waveforms $\phi_1$ as averaged over a 70-day period before each event as a reference and correlation waveforms $\phi_2$ obtained over a series of 5 day spans after the events. The relative dilation across the events are the same as seen in figure~\ref{fig:PFdvv} (top), of order  $5 \times 10^{-4}$. 
The values of $X$ for these pairs of waveforms varied between 60\% and 70\%. According to equation~(\ref{eq:PFerr}) divided by $\sqrt{78}$,  the value of $X$ would have had to be below 33\% if this large and apparent dilation were to be due to a random function with no actual dilation.  
The relative dilation between correlation waveforms before and after the event is therefore due to changes in seismic Green's function, and not to changes in the source of the waves.

\section{Summary}

Waveforms constructed by noise correlation can be extraordinarily sensitive to changes in material properties. Such waveforms are in principle affected by both changes in noise sources and changes in the acoustic properties of the medium in which the waves propagate. It has been shown here that long-duration diffuse waveforms permit changes in the source of the noise to be distinguished with high precision from changes due to a temporal dilation.

An expression was derived for the \textit{rms} of the apparent dilation $\epsilon$ measured on two waveforms, when there is no actual dilation between the two. This apparent dilation can be an effect of \textit{e.g.} a change in noise sources. The \textit{rms} value thus allows us to distinguish between an erroneous dilation measurement due to waveform change, and a physical wavespeed change in the medium. 

We have tested the validity of the \textit{rms} value using data from laboratory experiments, and we find that the theory predicts errors well. 

\begin{acknowledgments}
Most of this work was done while the first author was a visitor at the Laboratoire de G\'eophysique Interne et Tectonophysique (Universit\'e J. Fourier). This work was partially funded by the ANR  JCJC08 {\it SISDIF} grant, and by the ERC \emph{Whisper} Advanced grant. All the seismic data used in this study came from the Parkfield HRSN and were collected by the Berkeley Seismological Laboratory (BSL).
\end{acknowledgments}

\bibliographystyle{gji}

\label{lastpage}
\newpage

\end{document}